\documentclass[12pt]{article}
\begin{document}
\baselineskip=30pt
\title{
 3D Solution of Hartree-Fock-Bogoliubov Equations for Drip-Line Nuclei
 }
\author{
 J.~Terasaki, P.-H.~Heenen\thanks{Directeur de Recherches FNRS.} \\
 Service de Physique Nucl\'eaire Th\'eorique,\\
 U.L.B.-C.P.229, B-1050 Brussels, Belgium
 \and
 H.~Flocard \\
 Division de Physique Th\'eorique\thanks{Unit\'e de recherches des
 Universit\'es Paris XI et Paris VI associ\'ee au CNRS.},
 Institut de Physique Nucl\'eaire, \\
 91406 Orsay Cedex, France
 \and
 P.~Bonche \\
SPhT
\thanks{Laboratoire de la DSM} -CE Saclay, 91191
 Gif sur Yvette Cedex, France
 }
\maketitle

\begin{abstract}
We investigate the possibility of describing triaxial quadrupole deformations
for nuclei close to the two-neutron drip line
by the Hartree-Fock Bogoliubov method taking into account
resonances in the continuum. We use a Skyrme interaction
to describe the Hartee-Fock hamiltonian and
a density-dependent zero-range interaction
to evaluate the pairing field.
The mean-field equations are solved in a three-dimensional
cubic mesh. We study the stability of the two-neutron separation energies
and of the description of the nuclear surface
as a function of the number of active mean-field orbitals
and of the size of the mesh.
The even Ni isotopes are used as a test case
and the accuracy as a function of quadrupole
deformation is studied by performing constrained calculations.
A  first application to the study of
the two-neutron separation energies in
Ni isotopes up to the drip line is presented.
\end{abstract}

\newpage
\section{ Introduction }

Radioactive beam facilities, whether already active or soon to come,
will enlarge our knowledge of the nuclear structure far away
from the stability valley~\cite{DV89,MS93}.
They will  allow a complete investigation of the drip
lines for the light elements.
Beyond their importance  for
models aiming at
describing catastrophic astrophysical processes, such experiments
give the
possibility to address questions related to our understanding of
single-particle
or quasi-particle mean-field approximations.
One of these  questions concern  the isospin properties of the nuclear
interaction~\cite{CHA95} which plays a major role
in determining  the location of the drip lines.
Another one~\cite{DLM75,DHN94} is related to
the stability of the shell effects and of the magic numbers
in the vicinity of the drip line. This
stability depends in part on the modifications of
the spatial structure of the single-particle potential and especially
of its surface.
Large values of $\mid N-Z \mid$
may also generate some decoupling between
the proton
and neutron mean-fields.
In that respect, a  still open  question is the existence of nuclei
with neutron and proton densities showing
different deformations~\cite{BFH85,WSN94}.

Several extensive studies of exotic nuclei using
macroscopic-microscopic
method{\bf s} are already available~\cite{MN88,MNM95,YPD92,BCD94}.
They make predictions for binding energies, radii and deformations.
 However these
methods assume a simple parametrization of the shape
of the mean-field and of its radial dependence.
 Thus, it is difficult
to take unambiguously into account effects related to a large
excess
of neutrons especially when they generate  exotic phenomena  such as halos
or neutron skins\cite{MS93}.
Most studies run into difficulties because, in the vicinity of the drip line,
the Strutinsky method requires that one takes into account resonant single
particle states \cite{NWD94}.
Microscopic methods such as Hartree-Fock (HF) or Hartree-Fock-Bogoliubov
(HFB) which let the mean-field adjust self-consistently to the dynamical
conditions associated with a specific $N,Z$ configuration are better suited
for theses studies.

 The calculation of
nuclear properties in the vicinity of the drip
lines is made difficult
by the  very precise phenomenon which defines  them:
the Fermi level of one nucleon species becomes close to zero.
As the Fermi energy diminishes the active pairing space extends
into the continuum, and achieving a correct
description of pairing correlations becomes a major problem.
Simple methods such as HF+BCS are no longer valid because they predict
a leakage of nucleons into the continuum.
The  HFB method however provides a remedy to this deficiency
as it allows the
coupling to positive energy resonant states while preserving confinement
of the nuclear density~\cite{BU80,DFT84}.
 As a consequence, one must achieve an accurate description of the
resonances
which are immersed in a sea of continuum states.
This question has been successfully dealt with  in studies restricted
to spherical nuclei~\cite{DNW95}.

Static deformations near the neutron
drip line
have been observed for Sodium and Magnesium isotopes~\cite{DET79}
and are inferred for some Sulfur elements
from their $\beta$ decay properties~\cite{SOR93}.
Studies of these nuclei have been limited to shell model
calculations including two major shells~\cite{PM87,WBB90}
and to pure HF or relativistic Hartree calculations~\cite{WSN94}.
All these methods are of limited applicability for heavy nuclei,
either because the calculations become untractable or due
to the absence of a correct treatment of pairing correlations.
In  this
work we present a method which allows a solution of the
HFB equations in three dimensions and opens therefore the possibility of
an investigation of deformed neutron-rich nuclei.
The above described problems
are especially acute near the neutron drip line.
Indeed, in the vicinity
of the proton drip line, the existence of the coulomb barrier allows to
circumvent them  easily  at least for a determination of
global observables (energy, density distribution)
although not for more detailed
properties such as the lifetime of proton  emitters.
For this reason, all the calculations presented in this work will concern
the neutron-rich side of the nuclear chart.

Our method is similar to that
developped to study rotating
nuclei~\cite {GBD94,THB95}.
The next section briefly recalls some characteristics of the method.
It presents the modifications necessary
for an investigation of
neutron-rich nuclei and defines the parameter which control the numerical
accuracy.
In  Sect.~3
 we discuss the numerical tests which establish the quality of
our method.
Sect.~4 gives a first application to a series of Ni isotopes.
The last section collects our conclusions.

\section{ The Method }
We have developed a method to solve the HFB
equations by discretization  of a rectangular box~\cite{BFH85}.
This
has the advantage over methods relying on expansion on a basis
that nuclei with much
different deformations can be treated at the same level of accuracy.
In this section, we introduce our notations, paying special attention to
the points which
require some particular
care in the treatment
of nuclei far from stability.

Let us start by a brief reminder of the HFB equations for time-reversal
invariant states.
We write them as a function of  basis states {$i,{\bar i}$} related by a time
reversal operation.
The
HFB equation is
\begin{equation}\label{e2020}
\left(
\begin{array}{cc}
h-\lambda    &  \Delta \\
\Delta^\ast &  -h+\lambda
\end{array}
\right)
\left(
\begin{array}{c}
U \\
V
\end{array}
\right)
=
\left(
\begin{array}{c}
U \\
V
\end{array}
\right)
E_l \ ,
\end{equation}
where $\lambda$ is the chemical potential,
$h$ the mean-field hamiltonian:
\begin{equation}\label{e2030}
h_{ij}= T_{ij} + \sum_{kl}( V_{ikjl} + V_{i{\bar k}j{\bar l}})\rho_{kl}\quad,
\end{equation}
and $\Delta$ the pairing gap matrix{\bf :}
\begin{equation}\label{e2040}
\Delta_{ij}=\sum_{kl} V_{i{\bar\jmath}{\bar k} l} \kappa_{{\bar k} l}
\quad.
\end{equation}
In these expressions,
the one-body density matrix $\rho$ and
the pairing tensor $\kappa$  are defined
by:
\begin{equation}\label{e2050}
\rho       =  \left( V V^+ \right)\quad,\quad
\kappa     =  \left( U V^+ \right)\quad.
\end{equation}

In order to simplify the numerical resolution of
the HFB equation  and to make easier an adjustment of the effective
nucleon-nucleon force from which the matrix elements $V$
are calculated, we decided
to use two different forces, one acting in the particle-hole (ph) channel
and the other in the particle-particle (pp) channel.
The mean-field hamiltonian
$h$ is calculated according to eq.~(\ref{e2030})
with a Skyrme force, the
numerical results presented below have been obtained
with the parametrization SIII~\cite{BFG75}.
In
the pp-channel, we evaluate the pairing gap matrix $\Delta$~(\ref{e2040})
by means of a zero-range
density-dependent pairing force~\cite{THB95}:
\begin{equation}\label{e2060}
V_{\rm P}({\bf r}_1,{\bf r}_2) = V_0\,(1-P_\sigma)\,(1-\frac{\rho({\bf r}_1)}
{\rho_{\rm c}})\,\delta({\bf r}_1-{\bf r}_2)\quad .
\end{equation}
In the definition of $V_{\rm P}$, $P_\sigma$ is
the spin exchange operator and
$\rho({\bf r})$
the total density in coordinate space.
The pairing strength is controlled by $V_0$ while
$\rho_{\rm c}$ determines the $\bf r$-dependence
of the pairing gap field.

In a  previous study~\cite {THB95}, we have observed that a pairing field
acting over the entire volume of the nucleus yields a poorer reproduction
of moments of inertia than a surface peaked pairing field.
Since surface properties may play also a strong role for neutron rich
nuclei, we choose
a surface-peaked pairing field.
Within the parametrization~(\ref{e2060}) this is achieved by
selecting a value of $\rho_c$ close to the nuclear saturation density.
As this paper is devoted to the analysis of the convergence properties
of our method, we have differed an investigation of
the influence of the values of
$V_0$ and $\rho_{\rm c}$ on the properties of nuclei far
from stability.For this reason, we have chosen to
use the values  given
in ref~\cite{NWD94} ($V_0=1128.75 MeV~fm^{3}$
and $\rho_{\rm c}=0.13354 fm^{-3}$).
In fact, we have also checked that the numerical
properties of our method are
not affected by an increase of 20\% of these values.
Finally our pairing force is determined by a third parameter:
an energy-cutoff $\Lambda$ which controls the energy width of the pairing
active space.
In this work, this space extends up to 5~MeV above the Fermi
surface.
Since we are concerned with nuclei whose neutron chemical potential $\lambda$
is negative but close to zero, our HFB calculation takes therefore into
account resonances up to 5 MeV in the continuum.
Technically speaking, one must introduce $\Lambda$ to guarantee the existence
of a solution to the HFB equations with a zero range pairing force such
as $V_{\rm P}$.
The cutoff can be interpreted as a crude way of simulating the introduction
of a finite range which is known to ensure the convergence of the HFB
equations.
In order to select the magnitude of $\Lambda$, it is useful to remember the
physical motivation behind the HFB formalism.
It aims at describing the action of pairing correlations {\bf\it in the
vicinity} of the Fermi surface.
The energy scale typical of those correlations is given by the magnitude
of the pairing gap.
For nuclei, it is experimentally known to be of the order of 1~MeV.
We therefore think that the value of $\Lambda$ adopted in this work (5~MeV)
is sufficiently large to preserve the physically important features of the
HFB approximation while eliminating the divergence resulting from the high
relative momentum behavior of the matrix elements of a zero-range force.
Our
choice for the cutoff eliminates  a possible
coupling of
HFB eigenstates associated with deep hole states to
the eigenstates of the continuum.
As discussed in ref.~\cite{BU80, BST87},
such a coupling generates a width for deep
hole states.
As a matter of
fact, this phenomenon is by no means specific of nuclei far from
stability, however it
is generally neglected.
Besides,
the physical interpretation of such
widths, as well as their observation, is presumably very difficult,
especially if they turn out to be smaller than those resulting from
other broadening mechanisms~\cite{BBB83}.

Because the wave-functions of time reversed orbitals are related
to each other in a simple way, we only have to consider
the member $i$ of each pair ($i,{\bar\imath}$).
 From now on,
when we  will mention a number of orbitals,  that will be
the number of $i$-orbitals.
Moreover, because in the 3D calculations presented below we impose three
planes of symmetries for the nuclear density, we only have to store the
values of the single-particle wave-functions in one  spatial
octant.

The
HFB equation are solved in coordinate space by means of the two-basis
method described in ref.~\cite{GBD94}.
The coordinate space technique allows
to work with a limited number of basis states
however large enough to ensure
a good level of accuracy.
In particular our method describes equally well the interior and the surface
of the nucleus as long as the calculations are performed in a box
sufficiently large to enclose the nucleus and the tails of both the density
and the pairing tensor.
This point will be discussed below.
The method of solution relies on two nested loops of iterations.
In the outer one,
the eigenstates of $h$ are determined by the imaginary time-step
method~\cite{DFK80}.
In the inner loop
the HFB equations are solved, which gives
$U$, $V$, $\rho$ and $\kappa$
according to eqs.~ (\ref{e2020}-\ref{e2050}).
It is also in this loop that the value of $\lambda$ is determined by a
constraint on the particle number.

We denote by $\phi_i({\bf r},\sigma)$ the single-particle component of
spin  $\sigma$ of the i$^{\rm th}$eigenstate of $h$ and $e_i$ its
eigenvalue.
 On
this basis, refered
to as the HF basis, the  pairing operator $\Delta_{ij}$ is given by:
\begin{equation}\label{e2070}
\Delta_{ij} = \int d{\bf r} \Psi_{ij}^\ast({\bf r})
\widetilde\Delta({\bf r})\ ,
\end{equation}
where
\begin{equation}\label{e2080}
\Psi_{ij}({\bf r})=
\sum_\sigma \phi_i({\bf r},\sigma)
 \phi_{j}({\bf r},\sigma)^*\quad,\quad
\widetilde\Delta({\bf r})=V_0 ( 1 - \frac{\rho({\bf r})}{\rho_{\rm c}})
 \sum_{kl} \Psi_{kl}({\bf r}) \kappa_{k l}\quad.
\end{equation}
The function
$\widetilde\Delta({\bf r})$
is the local pairing gap in the coordinate representation.

The second basis, usually
called the canonical basis, is built from
the eigenvectors $\varphi_i$ of the density matrix $\rho$:
\begin{equation}\label{e2090}
(W^{\dag} \rho W)_{ij} = n_i \delta_{ij}\quad,\quad
\varphi_i = \sum_j W^t_{ij} \phi_j\quad .
\end{equation}
The eigenvalues $n_i$ are
interpreted as occupation probabilities.
In particular, the local density in coordinate space is given by
\begin{equation}\label{e2100}
\rho({\bf r}) = \sum_{i\,\sigma} n_i | \varphi_i({\bf r,\sigma}) |^{2}\ ,
\end{equation}
The feedback of the internal loop on to the external one is achieved
via the dependence of $h$ on $\rho({\bf r})$ and on other densities.
A characteristic of our numerical method is the imaginary time-step
algorithm, which
enables us to carry {\it only} the eigenstates of  $h$
for the physics that we want to study, even though
the dimension of the basis states associated with our spatial
discretization
inside a cubix box is of the order of several ten  thousands.
In the  next section, we
especially study the convergence properties  of the numerical algorithm
with respect to the
number of eigenvectors in the HF {\bf or} 
canonical basis followed with the imaginary time-step method.

\section{Numerical Analysis of the Discretization of HFB equation}

Before studying a chain of isotopes, we discuss the influence of the
continuum on the HFB solution for the neutron rich nucleus $^{84}$Ni.
To solve the HFB equations, we impose that the density vanishes at
the edges of the box, the resulting boundary conditions imply that
the positive energy spectrum is discretized.
Thus we have to investigate the convergence of the results with respect to
the size of the box.

Our choice of $^{84}$Ni is motivated by two reasons.
First this nucleus has a large neutron excess with a
neutron Fermi energy
close to zero (-0.98 MeV).
Second, $Z=28$ is a well defined magic number
and proton pairing correlations vanish.
This simplifies our analysis as it
allows a study of protons within the Hartree-Fock approximation,
limiting thereby our HFB study to effects of the
continuum on neutron pairing only.
{}Furthermore, again
because of the proton magicity, the
ground state is spherical.
We can therefore solve the problem in a cube and investigate
the convergence
versus the only remaining numerical parameter of the problem:
namely the dimension $R$ of the cubic box.
Let us recall however that in our method of solution
of the mean-field equations, spherical configurations
are treated in the same way as axial and triaxial deformed ones.

\subsection{ Single-Particle Spectrum}

According to the  presentation of the two-basis method,
two different sets of orbitals are considered in HFB calculations:
the HF-basis (eigenstates of $h$ with eigenvalues  $e_i$) and the canonical
basis (eigenstates of $\rho$ with eigenvalues $n_i$).
These two sets, identical in the BCS limit, are not very different.
Since their global features are similar, we discuss
first the canonical basis.
We will later point out some of their differences
in the particuliar case of
resonant states.

In Figure~1, we have plotted the diagonal elements $h_{ii}$ of the mean-field
hamiltonian in the canonical basis as a function of the box size for the
states in the vicinity and above the Fermi level.
As expected, two different classes appear in the positive energy part of the
diagram.
First some are stable with respect to $R$,
they correspond to resonances.
For the others, that we will refer to as continuum states, $h_{ii}$ is a
steadily decreasing function of $R$.
Each
resonance possesses quantum numbers which allow an unambiguous assignment
to a single-particle shell model state.
In one case ($1h{11/2}$), the resonant states are crossed by continuum states
as $R$ grows. Note that due to the fact that spherical symmetry is not
imposed in our calculation, the degeneracy between the different subshells
is slightly removed.

Figure~2 shows that the occupation $n_i$ of the resonant states is small but
non negligible. It is also stable versus $R$.
Nevertheless, one observes a slight decrease of $n_i$.
This phenomenon can be correlated to the evolution of $h_{ii}$ for the
$3s_{1/2}$ bound state whose energy  decreases slowly as a function of $R$
(see figure~1).
Within the HFB formalism, this could reflect a coupling of this bound state
to the continuum.
On the other hand, since the effect is observed
for an $s$-state, it may also be due to the
improved description of the nuclear potential at large distance as the
box size increases.
In order to test the latter hypothesis, we have performed
HF calculations in boxes of growing sizes.
In this case, we have found that the
energy of the $3s_{1/2}$ particle state decreases by 400~keV
as the box size varies from 11 to 15 fm.
Therefore we can assert that improving
the description of the nuclear surface is the main cause for
the lowering of the $3s_{1/2}$ energy. In fact, pairing
correlations acts only as a second order effect by slightly
reducing the lowering.
 Within the HFB formalism, this lowering of the $3s_{1/2}$ energy
translates
into an increase of occupation probability
which can be observed in the upper part of figure~2.
As a consequence, because all occupations add up
to the neutron number $N$, the $n_i$'s of the
resonances diminish.
Figure~2 shows also that the occupation of all the continuum states
remain zero,
irrespective of the value of $h_{ii}$.
In particular, for the largest box sizes, some of these states are below
resonant states with non zero occupation.

 The main difference between the HF and canonical bases
is that resonant orbitals can be unambiguously identified
only in the latter one.  The mixing of resonant and
continuum states in the HF basis has many consequences.
First, the evolution versus $R$ of the diagonal matrix elements of $\rho$
in the HF basis is not as simple as the one shown on figure~2.
Accidents occur for values of $R$ such that a continuum
state crosses a resonance.
A study of the wave-functions for the resonant and
continuum states also shows this qualitative differences between
both bases.
 Let us consider for instance
the six resonant eigenstates  of $\rho$ for which $h_{ii}$ is constant
and close to 3.1 MeV when $R$ varies from 11 to 15 fm,
grouped under the symbol  [ on figure~1.
These eigenstates
can be assigned to the spherical $1h_{11/2}$ level.
In the canonical basis, their wave-functions $\varphi_i$ are
localized near the origin.
In that respect, they differ significantly from
continuum states with similar values of $h_{ii}$.
On the other hand, in the HF basis, when $R$ is between 13 to 15 fm,
there are
{\it eight} states with $e_i$ close to 3~MeV with wave-functions $\phi_i$
having a large amplitude near the origin.
These wave-functions exhibit also a tail at large distances, which
indicates that the decoupling between resonant states and continuum is
not realized in the HF basis.
As an example, we show in
the lower part of figure~3
the quantity  $\sum_\sigma\vert\varphi_i(x,y,0, \sigma)\vert^2$ for one of
the $1h_{11/2}$ states.
 In contrast, the
upper part shows $\sum_\sigma\vert\phi_i(x,y,0, \sigma)\vert^2$ for the
HF-orbital having the maximum overlap with this resonance.
The
latter function displays a tail beyond 9~fm that is not present for
the corresponding quantity in the canonical basis.
 This comparaison confirms that the canonical basis is the relevant one
for the analysis of the resonances.

\subsection{ Total Energy and Two-Neutron Separation Energy}

We turn now to global observables to study the stability of our results
as a function of the number of states available for the description of the
continuum.
Figure~4 shows the total energy of $^{84}$Ni as a function of $R$.
When the number of HF (or canonical) basis states is at least
equal to 60, convergence is achieved.
The $R$-converged value of the energy is determined with an
accuracy better than 100~keV when the box size is equal to 12~fm.
The behavior of the curve obtained with 55 neutron states can be understood
from figure~1.
 For
$R$ greater than 14~fm, the $1h_{11/2}$ resonance has been crossed by
some continuum states.
With 55 states only, one cannot account simultaneously for these
continuum states and the 6 states of the resonance.
The increase of $E$ reflects therefore the loss of pairing energy due
to some resonant states leaving the active pairing space.

As can be seen from figure~4, the larger the box, the more states one
needs to reproduce correctly the total energy, up to 80 states at least in a
17~fm box. Indeed, as the size of the box increases, one need more and more
states to describe correctly both the resonances and the continuum to which
they couple.
This is confirmed by figure~5 which gives
the two-neutron separation energies $S_{\rm 2n}(N,Z) = -E(N,Z) + E(N-2,Z)$.
Already for $R$=12 fm, this quantity, which determines the location
of the drip line, can be calculated with a precision better than~100keV,
when the basis includes at least 60 states.

{} Finally,
one must mention that the convergence of $E$ versus $R$ is better than that
achieved for its separate components (kinetic, potential,
spin-orbit, pairing), which is a usual feature of
a variational method such as HFB.
For instance, the variation of the pairing energy when $R$ increases from
11~fm to 15~fm is of the order of 0.5~MeV
when 70 basis states are used.

\subsection{ Density and Pairing Field }
Since the possible existence of halos or neutron skins motivates in
part our interest in neutron-rich nuclei, it is important to determine how
well we can calculate the tail of the nuclear density.
In figure~6
using a logarithmic scale, we have plotted the neutron density of $^{84}$Ni
for several box sizes.
{} For
$R$ larger than 14~fm, the density can be predicted with good accuracy over
4 decades (i.e. up to 13~fm).
The comparison  between the neutron and proton densities does show
the existence of a neutron skin in this neutron rich nucleus.

In figure~7,
we compare the radial distribution of the neutron pairing field
$\widetilde\Delta_{\rm n}$~(\ref{e2080}) with the neutron and total densities.
The pairing field extends further than the density by about 2~fermi.
This point has been discussed in  ref.~\cite{DFT84} where it was shown
that, due to contributions of HFB eigenstates with quasi-particle energy
larger than $\vert\lambda\vert$, the density decreases asymptotically faster
than the pairing tensor.
{}From a practical point of view, it
means that the box size must
be large enough to accomodate the spatial extension of
$\widetilde\Delta_{\rm n}$.
This indicates that there must be an adequation between the cut-off in
coordinate space introduced by the box and the parameters of the pairing
interaction.

Let us discuss the problems associated with the 3 parameters of the pairing
interaction (equ.~\ref{e2060}).
We have checked that an increase by 20\% of the value of the pairing strength
$V_0$ does not modify the quality of convergence with respect to $R$.
The critical density $\rho_c$ that we have used here is smaller than
in~\cite{THB95}, leading to a more surface peaked pairing field.
This value therefore amplifies the surface character of the pairing field and
the conclusions of the present work should not be affected by an adjustment
of this density factor.
The third parameter of the pairing interaction is its cut-off.
To check a consequence of the change of this cut-off, we have performed
calculations without any.
This is equivalent to introduce an implicit $R$ dependant cutoff
$\Lambda(R)$ determined by the number of basis states.
{}From figure~1
one sees that for 70 basis states, $\Lambda(R)$ decreases from a value larger
than 10~MeV for $R=11$~fm to about 6~MeV for $R=15$~fm.
With such a number of orbitals,
the sensitivity versus the box dimension is enhanced: the total energy $E$
varies by 0.3~MeV when $R$ increases from 12 to 15~fm.
This indicates that the larger the cut-off (i.e. the greater
$\Lambda$), the larger the required number of orbitals.

\subsection{ Deformation Energy Curve}

One of the questions we want to address in the future
is the influence of static deformations on the position of the drip line.
Because they are proton magic,
Ni isotopes are spherical.
To test the ability of our model
to describe deformed nuclei, we have enforced
deformations by means of a constraint on the quadrupole moment
for $^{92}$Ni.
We have chosen this nucleus, instead of $^{84}$Ni, as it is in our calculation
the last stable Ni isotopes against two-neutron emission before the drip line
(see next section). Figure~8
presents thus the axial deformation energy curve of $^{92}$Ni as a
function of the mass quadrupole moment.
In figure~9, we show that prolate and oblate excitation energies at
fixed quadrupole moments are stable versus the box size.
This demonstrates that we can calculate spherical and deformed HFB solutions
of neutron-rich nuclei with similar accuracy.

For a dimension of the basis equal to 70, we find that the convergence with
respect to $R$ of the absolute energies is not as good as it was
for $^{84}$Ni:
when $R$ grows from 12 to 15~fm, we observe a variation of 0.1~MeV.
Increasing the basis size to 80 mean-field wave-functions
brings a gain of energy of 50 to 100~keV for each box size and
decreases the amplitude of the variation with $R$ to 50~keV.
This enables us to estimate the errors on total energies
to be of the order of 100 to 200~keV with 70 wave-functions
for $^{92}$Ni, which is still sufficient for an accurate
description of two-neutron separation energies.

The results discussed in the last section have been obtained with a box size
equal to 15 fm and 70 basis states.
A smaller box is definitively not appropriate whereas a larger one
demands more states without any significant improvement of the
results.

\section{ Ni Isotopes two-neutron Separation Energies}
As a first application of our method, we have considered the Ni isotopic
chain, which is experimentally known up to $^{78}$Ni~\cite{ENG95}.
The two-neutron separation energies $S_{\rm 2n}$ are shown in figure~10.
The comparison with experiment~\cite{WA85} is satisfactory.
The two major accidents in the curve correspond to the magic numbers $N=28$
and 50.
The curve crosses zero at $A=94$ which is therefore our prediction for the
drip line associated with the emission of a neutron pair.
In figure~10,
we have also reported HFB spherical results of ref.~\cite{DNW95}
which have been
calculated with the SkP interaction~\cite{DFT84}.
The overall slope is similar and the predictions for the position of drip
line is the same.
 The SkP curve displays only very small slope changes at major shell
crossings.
This may be a consequence of the larger single-particle level
density for the SkP force than for the SIII force.
These different level densities are related to  the differences between the
effective mass values of SkP and SIII (1.0 and 0.75 repectively).
The smooth behavior of the SkP curve may also be due to stronger pairing
correlations.
This will have to be checked by more extensive studies.
On figure~10 the separation energy calculated
from the position of the Fermi level (i.e. $-2.0*\lambda_n$)
is also plotted.
It agrees nicely with the calculated S$_{2n}$, except around
the magic neutron number N=50.

Let us finally mention that we have investigated the stability of our
predictions for  $S_{\rm 2n}$ against the mesh size $\Delta x$ of our
calculation.
Decreasing $\Delta x$ from 1~fm to 0.75~fm results in a lowering of the
$S_{\rm 2n}$ by less than 50~keV over
the entire series of isotopes.
Such a small inacurracy does not modify our conclusions concerning
the position of the drip line.

\section{ Conclusion }

 In this work, we have presented a method of solution of the HFB equation
applicable to even nuclei in the vicinity of the drip lines.
It is well suited
to  the study of the neutron drip line where continuum effects can become
important especially when
the description of pairing correlations is crucial.
Our method is based on a discretization of the continuum in coordinate space
and is naturally adapted
to the investigation of
nuclei either spherical or with a static quadrupole deformation.
We have studied
the convergence of physically relevant quantities
(such as single-particle spectrum, binding energies and two-neutron separation
energies, neutron density distribution)
with respect to the dimension of the box in which calculations are performed.
We have shown that for a large but still tractable number of basis states,
we control this convergence, at least when the active range of the pairing
interaction does not extends further than 5~MeV into the continuum.
We have also shown how the size of the box is correlated with the number
of basis states and how to choose these two quantities to achieve a
reasonable convergence.

The method establishes a clean separation between positive energy resonant
states which contribute to pairing and continuum states which are mere
spectators of the dynamics.
Our analysis has confirmed that the canonical basis is the most convenient to
extract resonances out from the continuum.
We have also checked the expected behaviours of the density and of the
pairing gap field.
In particular we have verified that the latter quantity extends further out
than the density.
A motivation of the present work is to devise a method suitable to
investigate quadrupole deformation near the drip line.
Therefore we have tested that the accuracy of our method does not depend
on the deformation.
This has been done by an analysis of an energy curve obtained by means of a
constraint on the quadrupole moment.
However, the present analysis does not permit to investigate  bound
state widths
resulting from a coupling to the continuum.

As a first application we have calculated the two-neutron separation energies
for a series of Ni isotopes extending up to the neutron drip line.
Our results agree with data when available~\cite{WA85}.
We predict that the shell effect at $N=50$ remains strong and that the drip
line occurs at $N=66$, namely for $^{92}$Ni.
This is in agreement with the former prediction of ref~\cite{DNW95}
using the SkP interaction.

Are all nuclei along the drip line spherical?
We believe that we have in our hands
the tool  required to answer this question. An
analysis of the predictions of recent
effective interactions for quantities relevant for the
astrophysical r-process is another of the tasks that
we are now in position to carry out.

{\bf Acknowledgements}

We thank  J. Dobaczewski and W. Nazarewicz
for interesting discussions.
This work was supported in part by the contract PAI-P3-043
and by the ARC convention 93/98-166
of the Belgian Office for Scientific Policy.

\newpage

Figure Captions
\begin{itemize}
\item[Fig.\ 1] Diagonal matrix elements of the HF Hamiltonian in the
              neutron canonical basis for  $^{84}$Ni. The calculatation
              is performed with 70 neutron single-particle
               wave functions as a function of the box size $R$.
Solid and dashed curves denote positive and negative parity levels,
respectively.
The numbers of quasi degenerate levels corresponding to each resonance
are indicated in  paranthesis.
\item[Fig.\ 2] Eigenvalues of the density matrix
corresponding to the positive energy states and to the 3s$_{1/2}$  bound state.
\item[Fig.\ 3] Density probability of one of the
1h$_{11/2}$ neutron wave functions
calculated with $R$=15~fm in the HF (a) and canonical (b) basis.
\item[Fig.\ 4] Total energy $E$ of $^{84}$Ni as a function of $R$
obtained with a number of neutron wave functions varying from 55 to 80.
The origin of the energies is arbitrary.
\item[Fig.\ 5] Two-neutron separation energies $S_{\rm 2n}$ for $^{84}$Ni
as a function of $R$ and of the number of neutron wave-functions.
\item[Fig.\ 6]  Proton (dot-dashed-dashed) and neutron
densities along a radius r.
The dashe, solid and dot-dashed curves correspond to a calculation including
70 neutron wave-functions and $R$ = 13, 15 and 17 fm, respectively.
The dotted curve corresponds to $R_{\rm box}$ = 17 fm
and 80 neutron wave functions.
\item[Fig.\ 7] a) Neutron pairing gap $\widetilde\Delta_{\rm n}(x)$
(see eq.\ (\ref{e2090}))
for $^{84}$Ni ($R$ = 15 fm and 70 neutron wave functions)
as a function of the radius (spherical configuration).
b) Neutron (solid) and total (dashed) densities
as a function of $r$.
\item[Fig.\ 8] Potential energy curve of $^{92}$Ni
as a function of the axial quadrupole moment $q$
($R$ = 15 fm and 70 neutron wave functions).
\item[Fig.\ 9] Total energies $E$ of $^{92}$Ni as a function of $R$
for prolate (open circles, q=3~b) and oblate (triangles, q=-4~b)
configurations excited
by approximately 1~MeV. The spherical configuration is taken as
the zero for each value of $R$.
\item[Fig.\ 10] Two-neutron separation energies $S_{\rm 2n}$ of even
Ni isotopes as a function of the mass number $A$:
experimental data~\cite{WA85} (open circles), present work
(solid circles) and results of ref.~\cite{DNW95}
(open triangles). The separation energies calculated
from the position of the neutron Fermi level are also shown (crosses).

\end{itemize}
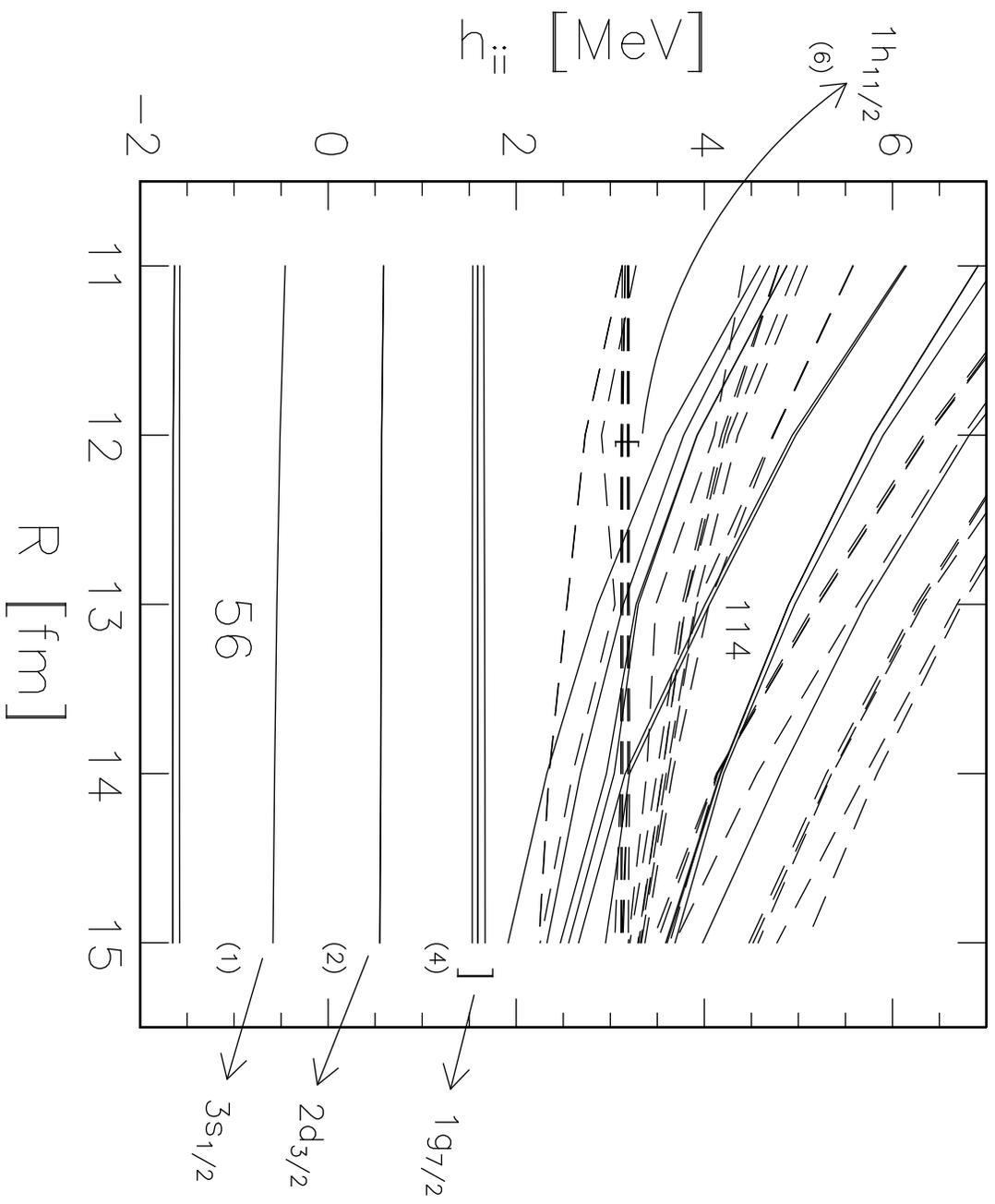
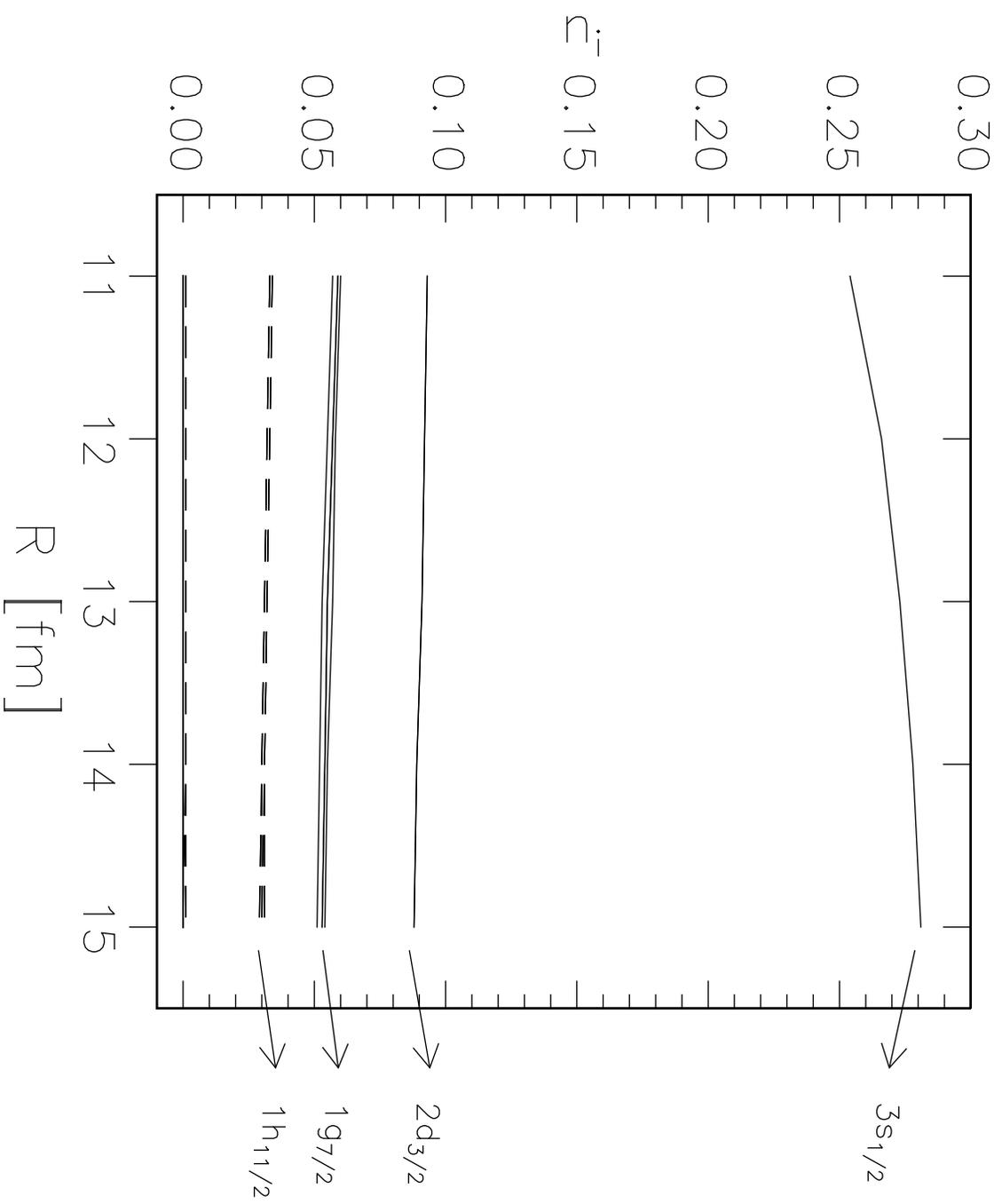
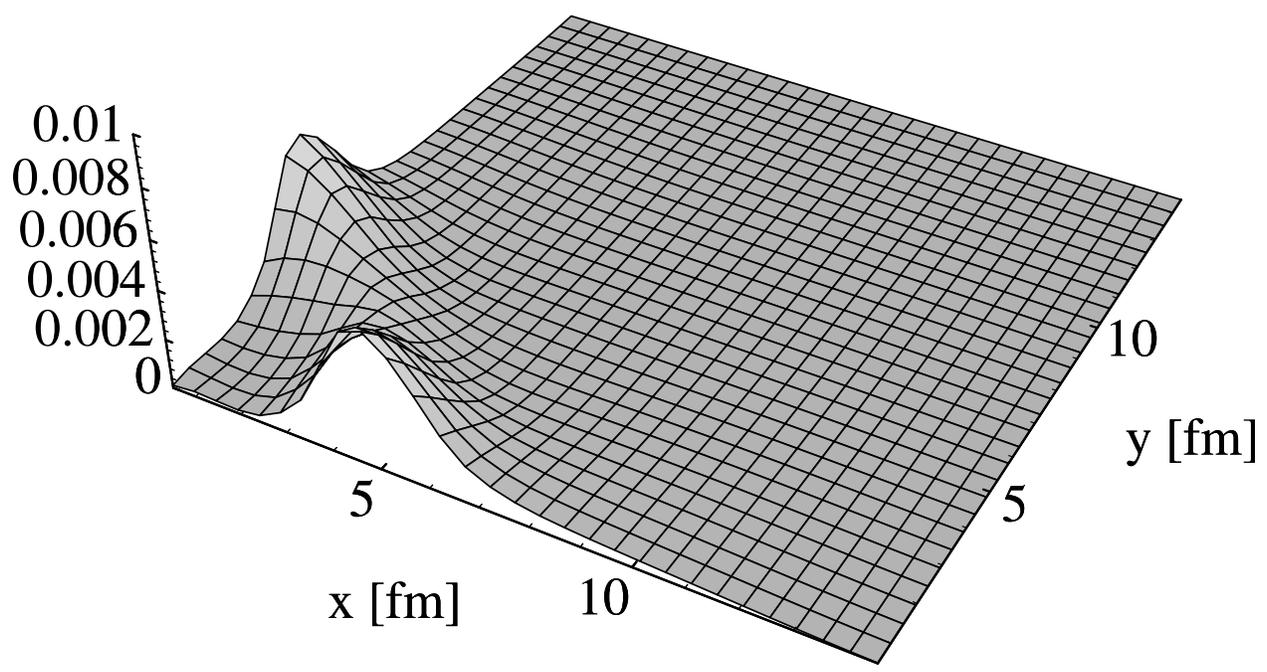
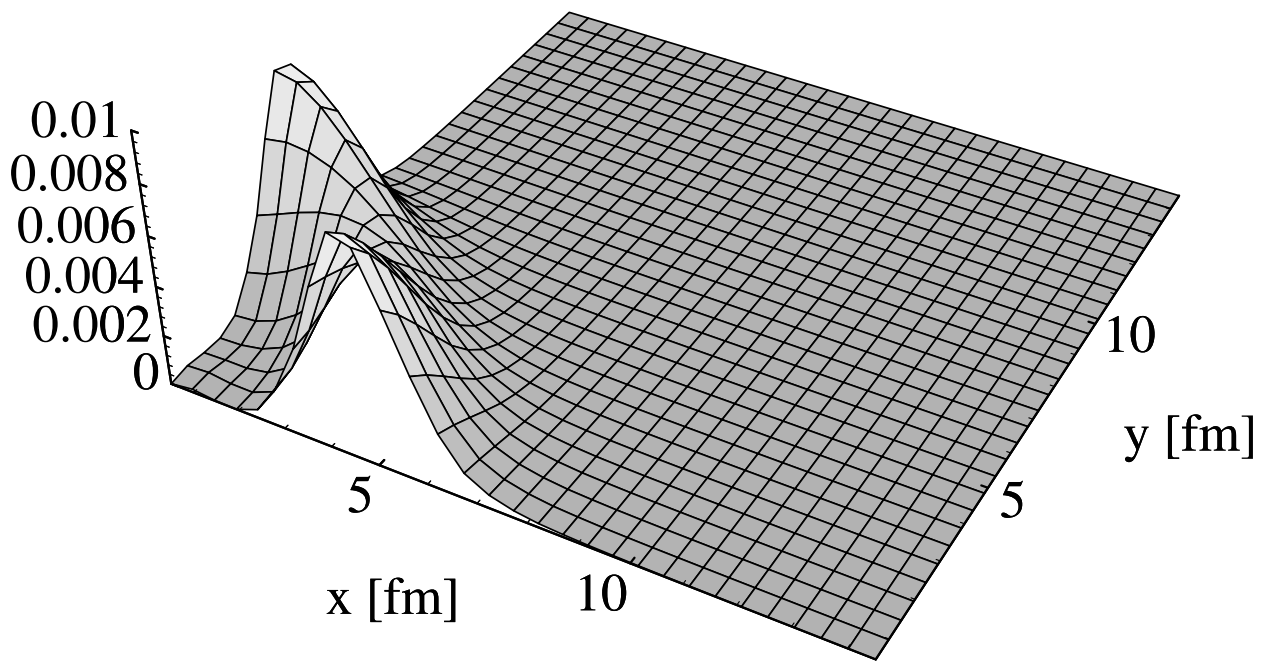
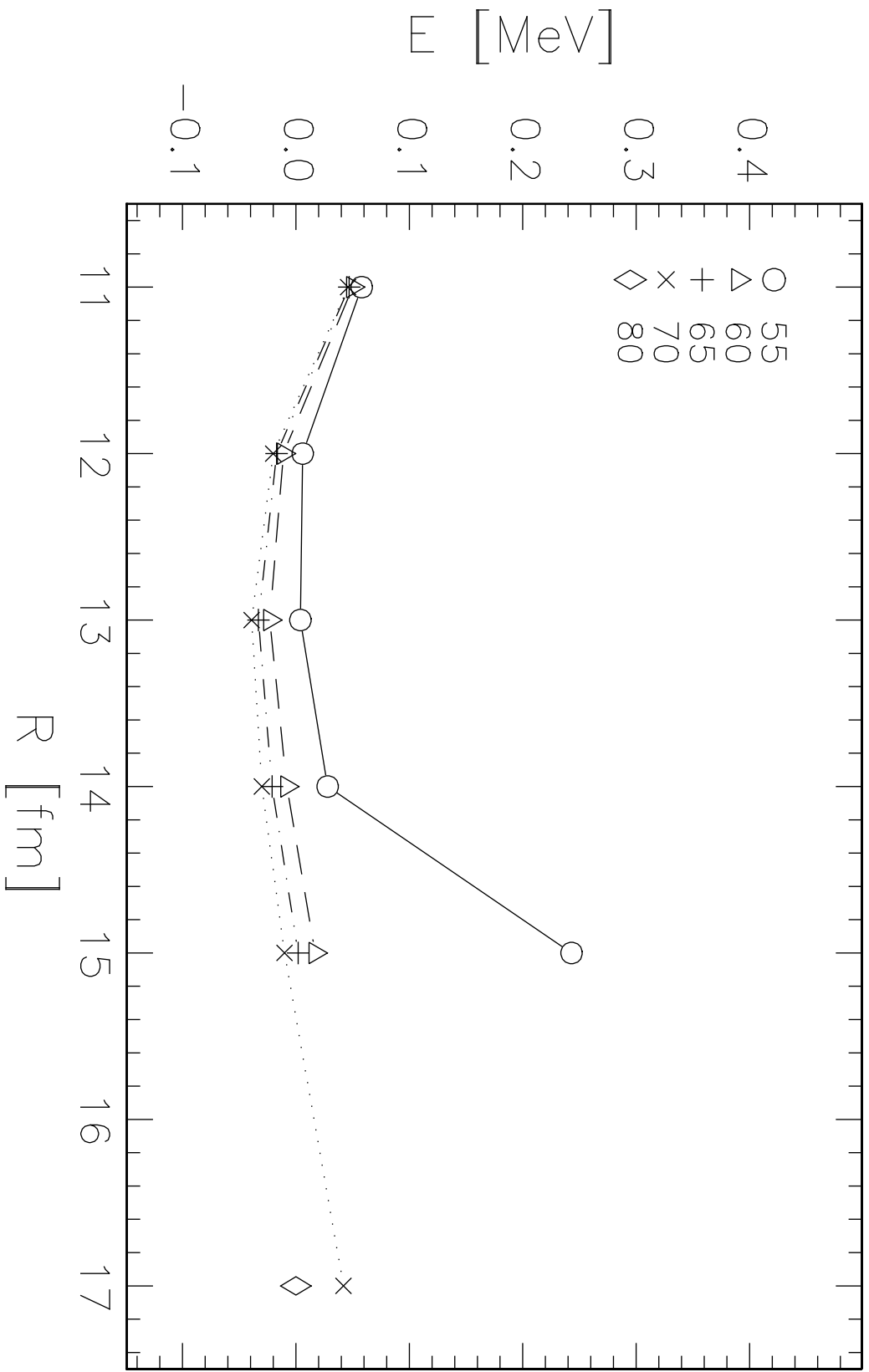
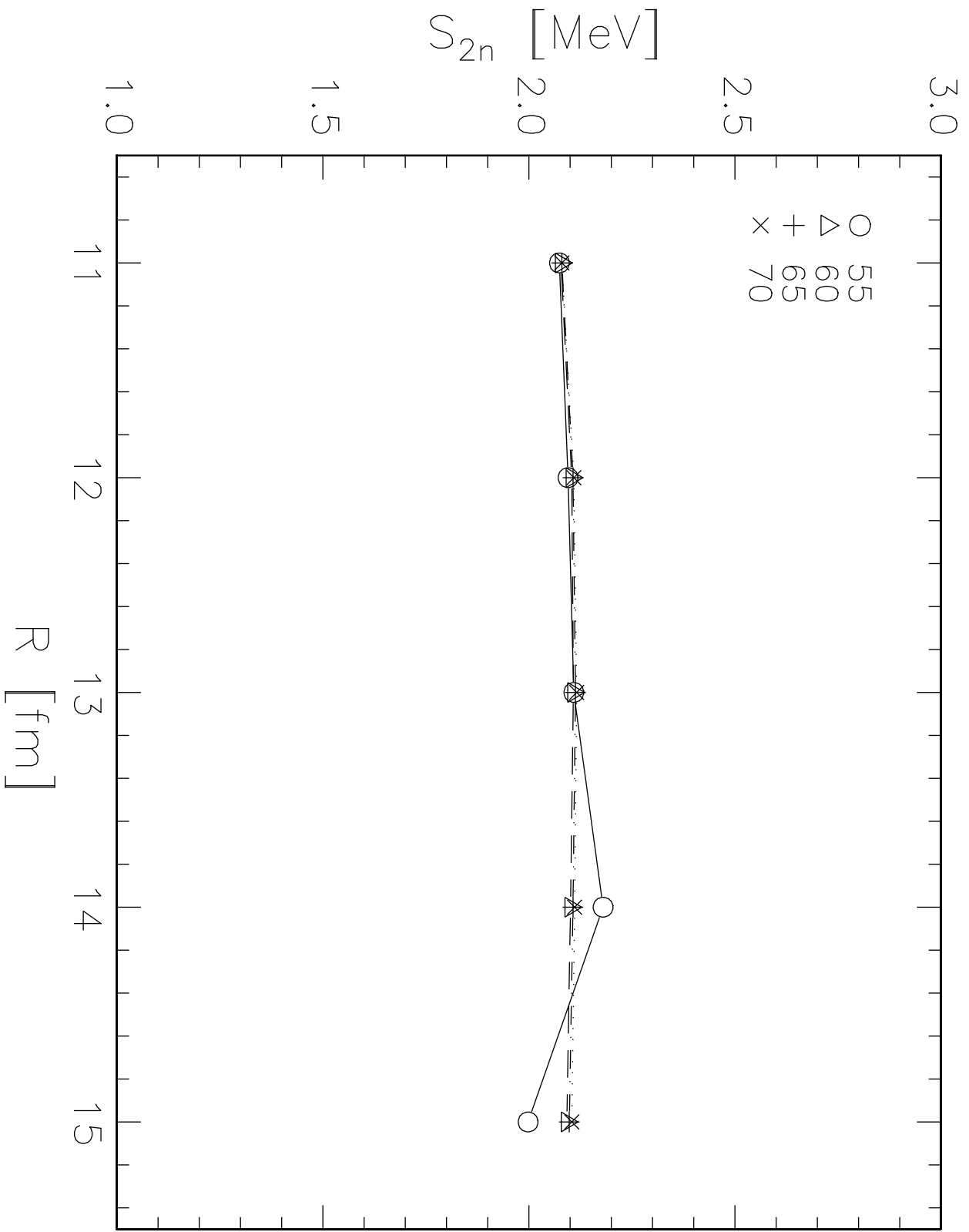
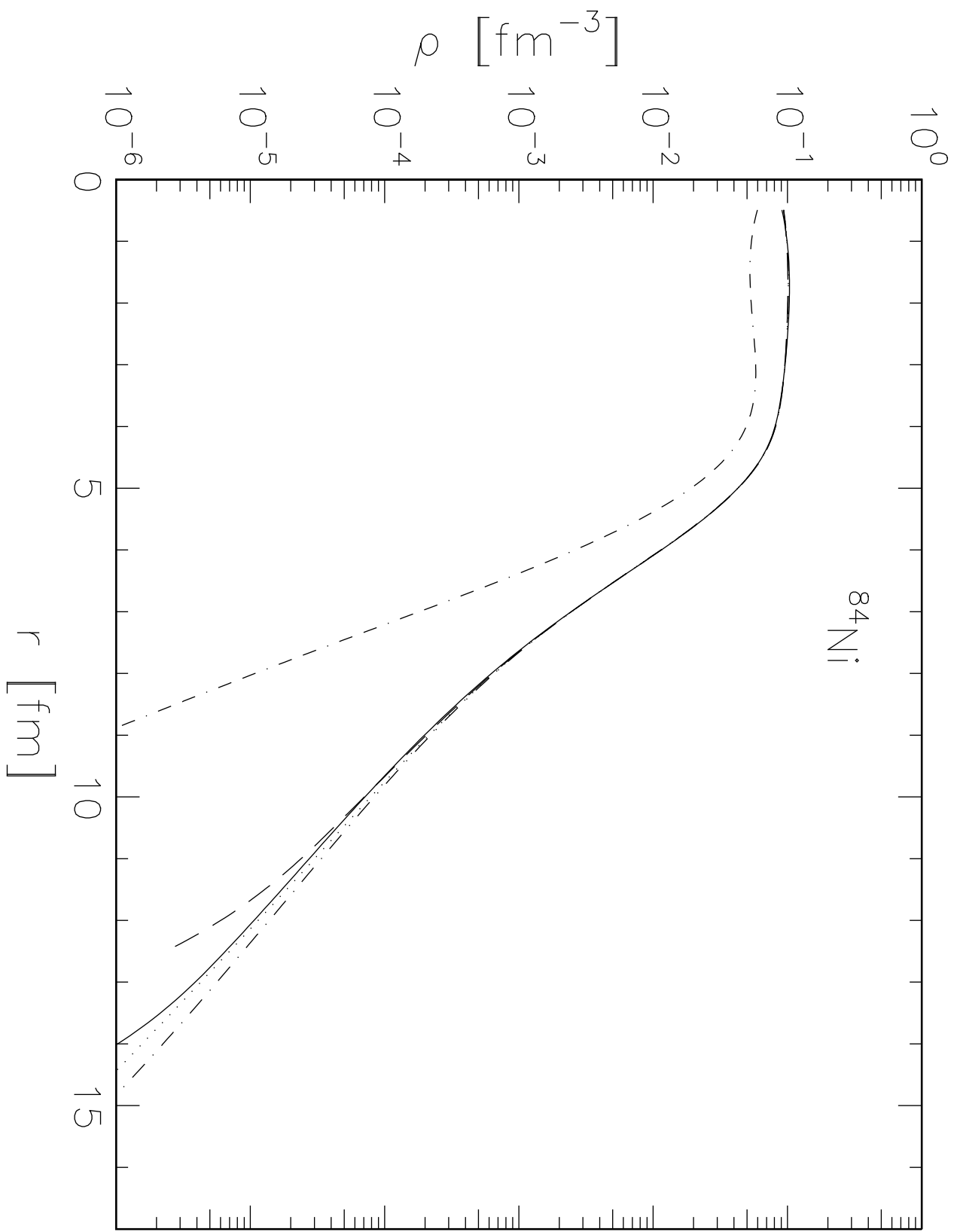
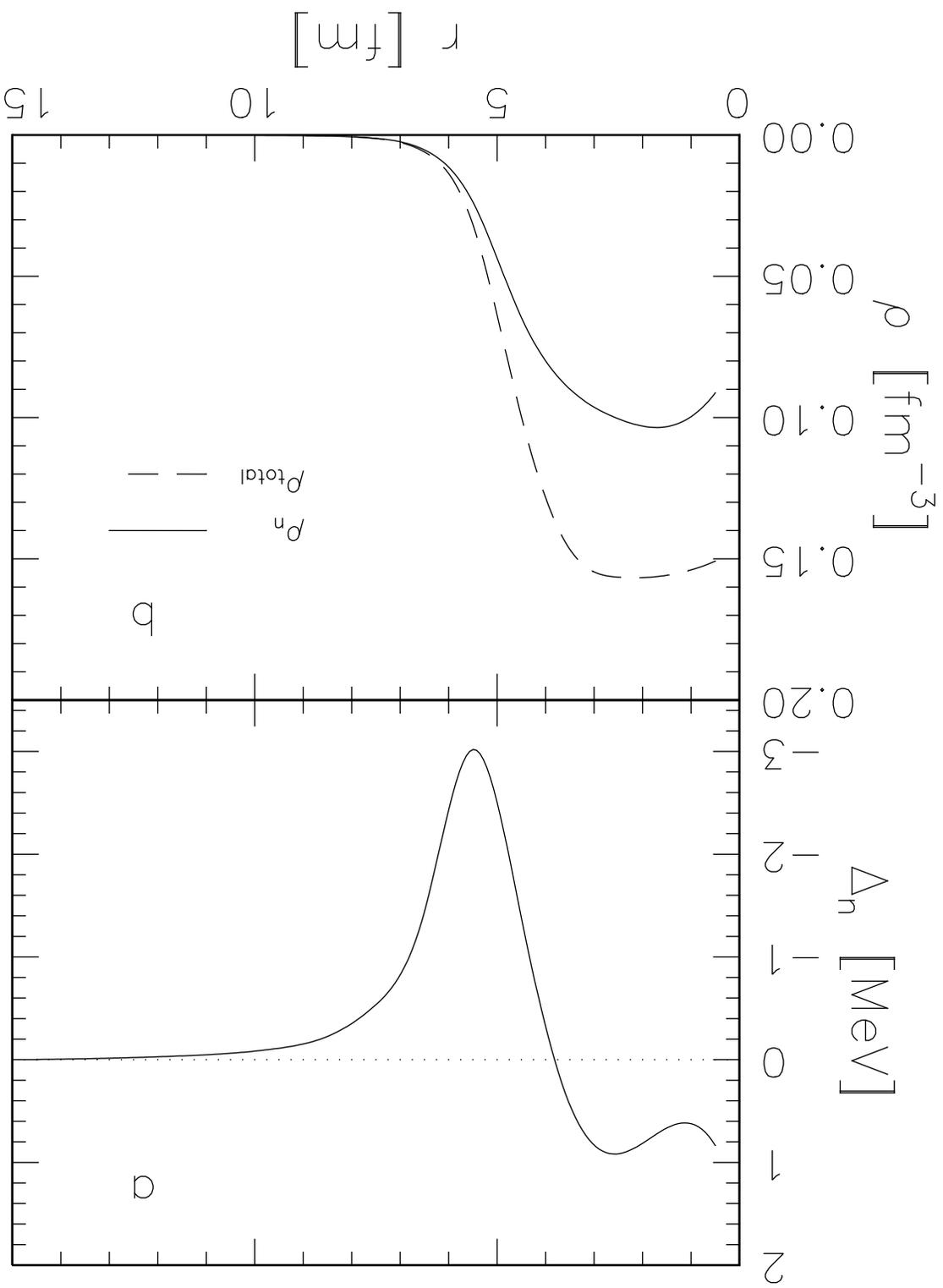
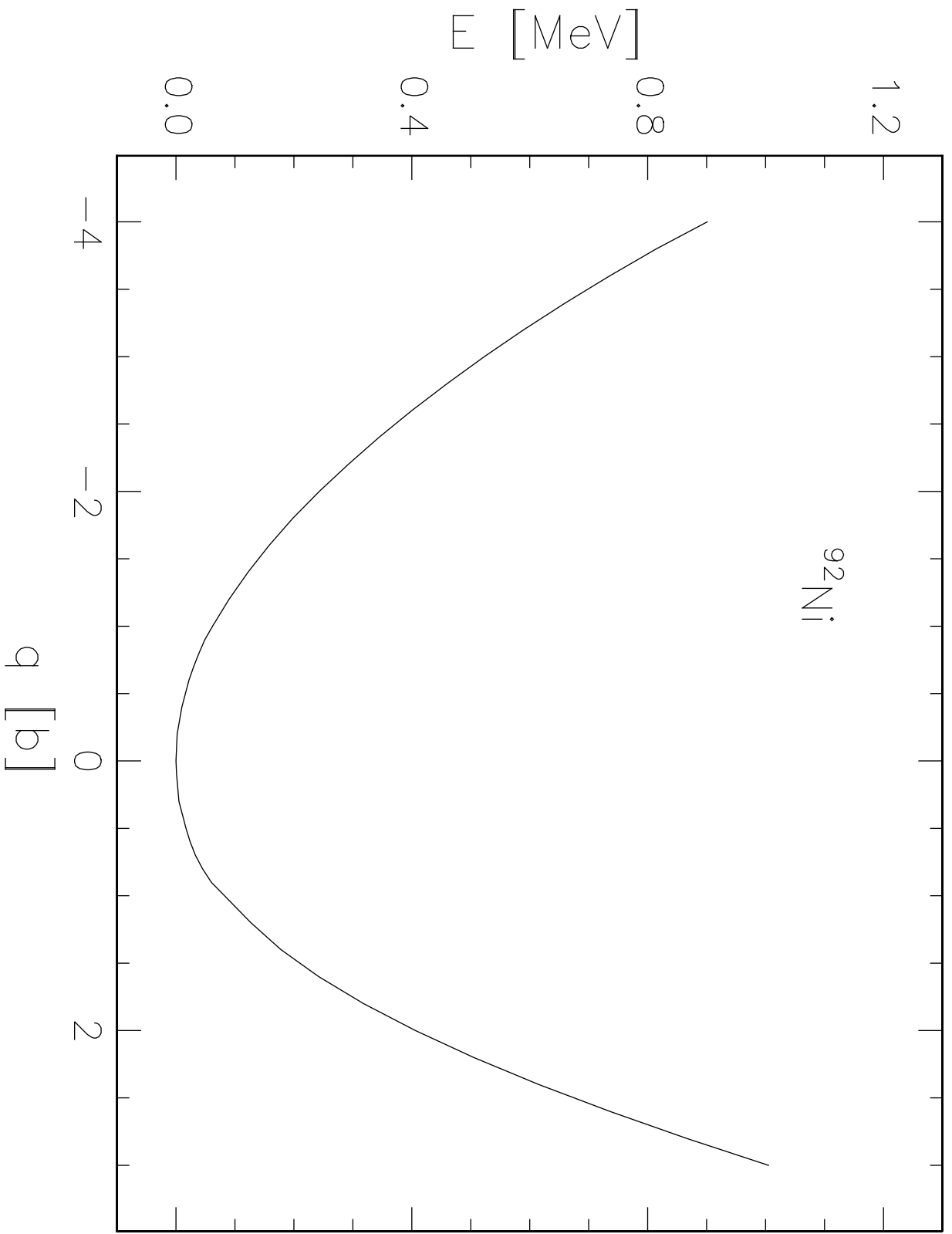
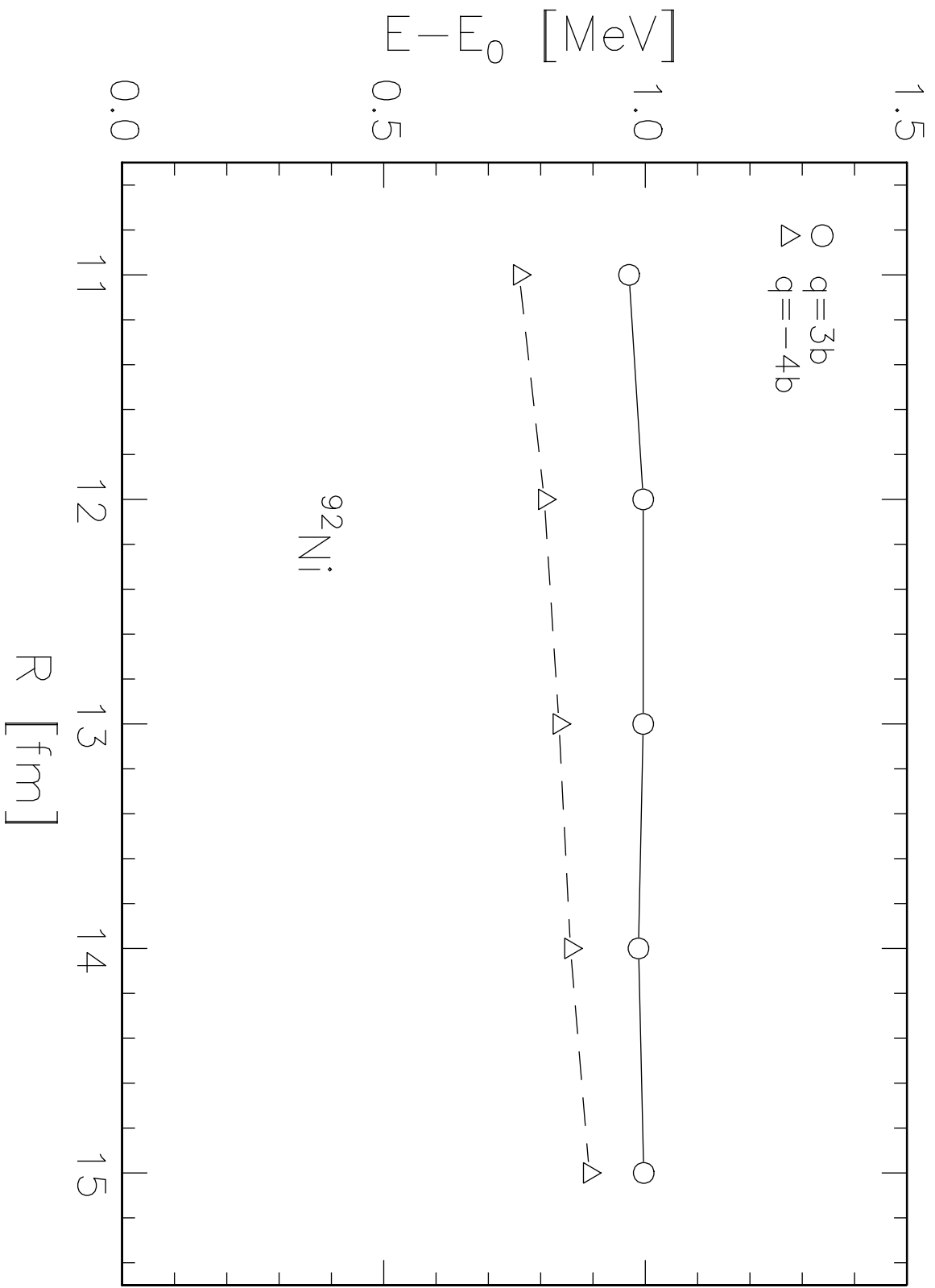
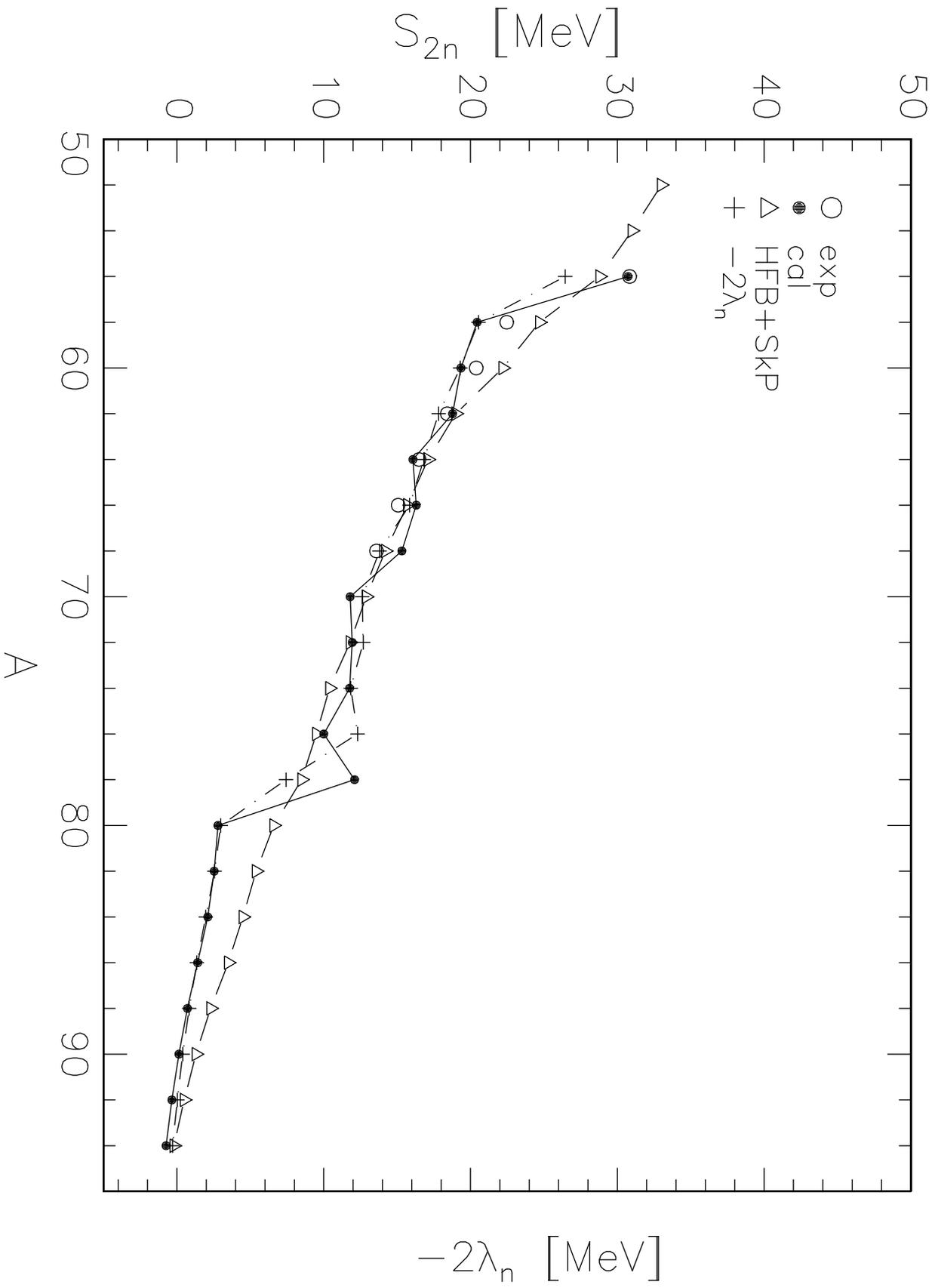
\end{document}